\documentstyle[prl,aps,multicol,epsf]{revtex}
\begin{document}
\title{Statistical Topography of Glassy Interfaces} 
\author{Chen Zeng}
\address{Department of Physics, Rutgers University, 
Piscataway, NJ 08855, USA} 
\author{Jan\'e Kondev} 
\address{Department of Physics, Brown University,
Providence, RI 02912, USA}
\author{D. McNamara and A. A. Middleton} 
\address{Department of Physics, Syracuse University,
Syracuse, NY 13210, USA}

\date{\today}
\maketitle  

\widetext

\begin{abstract}  
Statistical topography of two-dimensional interfaces in the presence of
quenched disorder is studied utilizing  combinatorial optimization 
algorithms. Finite-size scaling is used to measure geometrical 
exponents associated with contour loops and fully packed loops. 
We find that contour-loop exponents depend on the type of
disorder (periodic ``vs'' non-periodic) and they satisfy scaling 
relations characteristic of self-affine rough surfaces.
Fully packed loops on the other hand are unaffected by  
disorder with geometrical exponents that take on their pure
values. 
\end{abstract}


\begin{multicols}{2}
\narrowtext


Elastic manifolds in random media 
are used to model various condensed-matter systems with quenched disorder, 
including flux-line arrays in dirty type-II superconductors\cite{Blatter},
and charge density waves\cite{Gruner}. These disparate systems exhibit
a common feature, namely a low temperature glassy phase in which 
static and dynamic properties are dominated by the disorder.
One of the simplest models used to study this glassy phase 
is given by a 2-dimensional 
isotropic interfaces embedded in a 3-dimensional disordered medium.
The interface Hamiltonian in the continuum limit is
\begin{equation}
H( \{ h({\bf x}) \} ) = \int d^2{\bf x}
\left\lbrack
{K\over2} |\nabla h({\bf x})|^2 + V[{\bf x}, h({\bf x})]
\right\rbrack
\;\; ,
\label{eq1}
\end{equation}
where $h({\bf x})$ is the height of the interface above a 
flat basal plane indexed by a 2-dimensional vector
${\bf x}$, and $K$ is the  elastic stiffness
constant. Depending on whether the otherwise
uncorrelated random potential $V[{\bf x}, h({\bf x})]$ is
non-periodic or periodic 
along the height direction, the interface is usually referred to as a 
{\it random manifold} or a {\it random elastic medium}, 
respectively\cite{Balents}.

Much of the analytical understanding of these {\it glassy} interfaces
comes from functional renormalization group calculations\cite{RG1,RG2}.
The random manifold is found to have  a zero-temperature 
fixed point characterized by a  
roughness exponent 
$\zeta \approx 0.41$\cite{RG1}, i.e., the width ($W$) 
of the interface, as measured by the square root of the height variance, 
scales with its lateral linear size ($L$) as $W\sim L^{\zeta}$.  
The random elastic medium on the other hand exhibits a low-temperature fixed 
line below a glass transition. Along this line the interface is 
{\em super-rough} ($W\sim\ln(L)$),  
while in the high temperature phase it is 
{\em marginally rough} ($W\sim \sqrt{\ln(L)}$)\cite{RG2}. 
Recent numerical studies of the {\it exact} ground-state roughness 
based on combinatorial optimization algorithms strongly support 
the above picture\cite{GS-numerics}.

In this letter, we apply further refined implementations of these
polynomial-in-time algorithms to study the topography\cite{isich} 
of disordered 
interfaces at zero temperature. In contrast to the usual paradigm of a   
rugged landscape\cite{Sherrington} where the surface morphology is   
described in terms of height-height correlations, 
we instead focus on extended, non-local features of the random 
geometry as expressed by contour loops (i.e., lines of constant height) 
and fully-packed loops (defined below). We measure different {\em geometrical}
exponents for the two types of loops 
and check various scaling relations among them (see Eq.~(\ref{sc_rels})). 
This approach has proven to be very useful in characterizing 
the morphology of rough surfaces whenever the {\em complete} height profile is 
available\cite{CL-scaling}.

From the results obtained for contour loops our main conclusion is that
the glassy interfaces studied are self-affine and rough with roughness 
exponents in good agreement with theoretical findings. 
The study of fully packed loops on the other hand  addresses 
the question of the effect of  quenched disorder on critical fluctuations.
Namely, the {\em discrete} interface models we employ map to 
disordered fully  packed loop models. In the 
absence of disorder this loop  model is  critical and its  
exponents have been calculated exactly using Bethe ansatz and
Coulomb gas techniques\cite{FPL}. Unexpectedly, here we find that the 
values of the pure exponents are {\em unaffected} by disorder.    



\paragraph{Models and algorithms}
To simulate both random and random-periodic interfaces, 
we consider a simple cubic
lattice with random bond weights (energies) and its directed
(111)-interface. This interface, to be precise, is defined on
the dual lattice with  each elementary plaquette (a square) 
intersecting a single bond of the simple cubic lattice. 
The cost of such an interface is defined to be 
the sum of the weights of all 
the bonds that it cuts. In the case of the {\it random manifold}
the integer-valued weight of each bond is chosen independently and uniformly 
in the interval [0, 1000]. When 
simulating a {\it random elastic medium} the random weights are sampled in  
the same way with the important difference that the weights of all the bonds  
{\it directly} above one another along the (111) direction are set equal. 
In other words, the bond disorder is uncorrelated for the former  
and periodic along the height direction for the latter,  
reflecting the intrinsic periodicity of the
random elastic medium\cite{Balents}.


The problem of finding the ground state of an  interface
in the presence of quenched disorder is that of minimizing
its total energy cost. 
Employing max-flow-min-cut and minimum-cost perfect-matching algorithms, 
for the case of random disorder and random-periodic disorder respectively,
enables us to determine the {\it exact} 
ground state interface in time that  
grows only {\em polynomialy} with $L$.
This allows us to simulate relatively large 
system sizes for many disorder realizations.

The max-flow-min-cut algorithm interprets the simple cubic lattice
as a {\it directed} flow network. Each bond of the lattice is directed
along the (111) direction, and its random weight specifies the
maximum amount of flow that can be accommodated in this
bond. The algorithm finds the interface of minimum cost by searching
instead for the maximum flow (with flow conservation) that can be
sustained between the bottom and top layers with the interface 
sandwiched in between; see Fig.1. 
The algorithm works because the
interface of minimum cost is the {\it bottleneck} through which all
flow must pass.

\begin{figure}
  \begin{center}
    \leavevmode
    \epsfxsize=6cm
    \epsffile{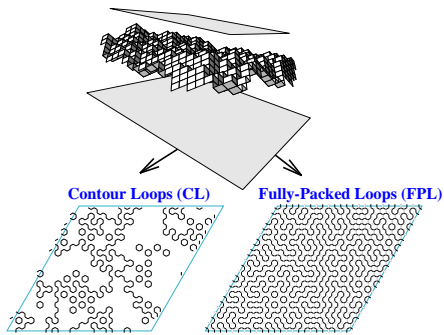}
  \end{center}
\caption{(111) interface of a simple cubic lattice
and its two loop representations. An example of
a ground-state interface confined between two flat
(111) layers is shown. The level set at mean height
consists  of closed contour loops due to periodic boundary
conditions. Also shown is the fully-packed-loop (FPL)
representation of the same interface (see text for details of construction).
}
\label{fig_111}
\end{figure}

Next we  note that the (111) projection of the
interface gives a rhombus tiling of the plane, which is also
equivalent to a complete (perfect) dimer covering of a hexagonal
lattice. The hexagonal lattice (${\cal L}$) in question is 
the dual of the triangular
lattice formed by the vertices of the rhombi, while the dimers 
lie perpendicular to their short diagonals\cite{TISOS}.  
This mapping is purely geometrical in nature, 
but when the disorder is periodic it facilitates 
the task of finding the interface of minimum cost. Namely, 
the periodic disorder can be 
represented entirely by random bonds on ${\cal L}$, 
and the minimization problem becomes one of finding
the perfect matching (dimer covering) with minimum cost. 
Detailed descriptions of both algorithms can be found 
in Ref.~\cite{Algorithms}. 


\paragraph{Geometric exponents and scaling}
Given the exact shape of the (111)-interface its topography is  
completely characterized by a contour plot with the  level spacing  
equal to a single step of the the discrete height. The contour plot 
consists of {\em contour loops} which live along the bonds of the 
hexagonal lattice ${\cal L}$. The contours are closed due to periodic 
boundary conditions which we impose in both lateral directions.  
For example, in Fig.1 we have drawn all the contour loops  
at mean height, for the (111) interface shown. The union of all the contour
loops for different realizations of disorder is the  {\em contour
loop ensemble}.

Apart from this natural contour-loop
characterization, there exists yet another interesting loop
representation of the (111)-interface, the fully-packed loops (FPL).
These loops owe their existence  to the 
one-to-one mapping between a (111)-interface and a
complete dimer covering of the hexagonal lattice ${\cal L}$. 
By removing the bonds of ${\cal L}$ that coincide with the dimers,
we are left with a configuration of fully-packed loops, as
shown in Fig.1, where every site of ${\cal L}$ 
belongs to one and only one loop. 
A physical realization of FPL are the magnetic domain walls in  
the ground state of the Ising antiferromagnet on the triangular
lattice\cite{TISOS}. FPL models of
general loop fugacity in the absence of disorder have been
studied recently\cite{FPL} and were shown to be critical for values of 
the loop fugacity that does not exceed two. The interface-FPL mapping thus 
allows us to consider the effect of quenched disorder on the critical 
FPL model on the honeycomb lattice with fugacity equal to one.

Following Ref.\cite{CL-scaling}, we consider the fractal dimension
of a loop $D$, the loop distribution 
exponent $\tau$, the loop correlation exponent $x_l$, and finally the 
interface roughness exponent $\zeta$.
These {\em geometrical} 
exponents are used to characterize the morphology of  
an interface which is
statistically invariant under the rescaling  $h({\bf x})\to b^{-\zeta}
h(b{\bf x})$, where $b>1$ is an arbitrary rescaling parameter; 
such interfaces are termed {\em self-affine}. 

The natural quantities associated with a loop are its length $s$ and the 
radius of gyration $R$, both  measured here in units of the lattice spacing.
For the ensemble of loops we define $n(s,R)$, the number density of
loops of length $s$ and radius $R$. Contour loops of a self-affine 
interface have no  characteristic length scale and we anticipate a 
scaling form for the number density:
\begin{equation}
\label{nsr_scaling}
n(s,R) = s^{-(1/D+\tau)} f(s/R^D) \ . 
\end{equation}
Integrating $n(s,R)$ over all radii gives the number density 
of loops of length $s$, 
\begin{equation}
\label{ns}
n(s) \sim s^{-\tau} \ , 
\label{tau}
\end{equation}
which we use to extract the exponents $\tau$ {\em and} $D$ (see the following
section).

Yet another measure that was introduced in Ref.\cite{CL-scaling} is  
the loop correlation function $g(r)$ which gives the  probability
that two points separated by distance $r$ belong to the same
loop. Just like in the case of the number density of loops we anticipate 
a power law 
\begin{equation}
g(r) \sim r^{-2x_l}
\ ,
\label{x_l}
\end{equation}
where $x_l$ is the loop correlation exponent. 

In Ref.\cite{CL-scaling} 
scaling relations were derived for the loop exponents 
$\tau$, $D$, and $x_l$ assuming a self-affine interface of roughness 
$\zeta$:
\begin{eqnarray}
\label{sc_rels}
D & = & 2 - x_l -\zeta/2 \nonumber \\
\tau & = & 1 + (2-\zeta)/D  \ .
\end{eqnarray}
These results follow from sum rules for the average loop length and 
the average number of {\em large} loops 
(i.e., those with radii comparable to $\rho$) inside an area of
linear size $\rho$. Eq.~\ref{sc_rels} tells us that 
there are only two independent exponents 
so measuring all four and checking the validity of the scaling relations
provides an important consistency check on the {\em assumed} 
self-affine nature of the rough interface. 
Furthermore, the second  scaling relation suggests a method for 
measuring the roughness exponent  directly from the loop data. 
Namely, integrating $n(s,R)$ over all $s$ and making use of 
$D(\tau-1)=2-\zeta$, we find
\begin{equation}
n(R) \sim R^{\zeta-3} 
\label{zeta} 
\end{equation}
for the number density of loops of radius $R$. 


\paragraph{Numerical results}
We first describe our results for the {\it 
random elastic medium}. Four different sample sizes, $L=72, 120, 240$,
and $480$ were simulated with 
$10^4$ disorder realizations for each size.

\begin{figure}
  \begin{center}
    \leavevmode
    \epsfxsize=7cm
    \epsffile{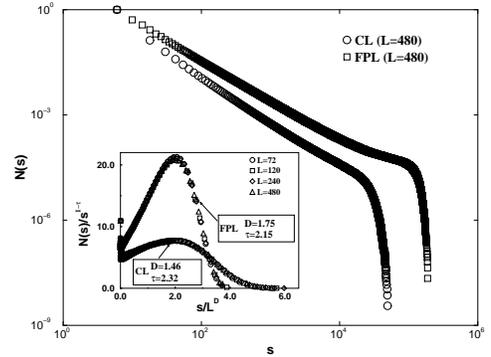}
  \end{center}
\caption{
Cumulative number density $N(s)$ (normalized by total loop number),
for CL and FPL;  $L=480$.
FSS analysis for $L=72, 120, 240$, and $480$, is shown in the inset
and used to determine exponents $D$ and $\tau$ (see text). 
$N(s)$ is binned in intervals of $0.08s$ at
successive $s$ to estimate the systematic errors which
are smaller than the symbols shown.
}
\label{fig_tau}
\end{figure}

The fractal dimension $D$ and loop exponent $\tau$ can be 
simultaneously extracted from $n(s)$ using a 
finite-size scaling (FSS) form $n(s) = s^{-\tau} f_n(s/L^{D})$,
which follows from Eq.~\ref{tau}. In order to minimize the statistical 
noise at large $s$, we consider instead the cumulative number density
$N(s)\equiv \int_{\tilde{s}>s} n(\tilde{s}) d\tilde{s}$,
with the scaling form $N(s) = s^{1-\tau} f_N(s/L^{D})$. 
Results are summarized in Fig~\ref{fig_tau}. 
The power-law scaling regime is evident over three decades
in loop length ($40<s<40000$). 
The deviations from scaling at small and large
$s$ are attributed to the lattice cutoff
and finite lattice size, respectively. 
The best data collapses yield
$D=1.46\pm 0.01$ and $\tau=2.32\pm 0.01$ for CL,
and $D=1.75\pm 0.01$ and $\tau=2.15\pm 0.01$ for FPL. 

\begin{figure}
  \begin{center}
    \leavevmode
    \epsfxsize=7cm
    \epsffile{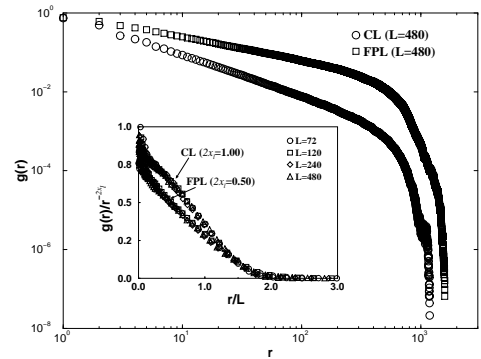}
  \end{center}
\caption{
Loop correlation function $g(r)$
for both CL and FPL; $L=480$. FSS plots are shown in the inset; systematic
errors estimated as in  Fig.~\ref{fig_tau} are smaller then the symbols shown.
}
\label{fig_x}
\end{figure}

We have also computed the loop correlation function $g(r)$, from
which the exponent $x_l$ is extracted using a
scaling form $g(r) = r^{-2x_l} f_g(r/L)$. The best data collapses,
shown in the inset of Fig.~\ref{fig_x}, yield $x_l= 0.50\pm 0.01$
and $x_l= 0.25\pm 0.01$ for CL and FPL, respectively.
An exact value ($x_l^{{\rm exact}}=1/2$) of the loop correlation
exponent for CL was proposed for self-affine rough surfaces 
{\em independent} of the roughness\cite{CL-scaling}. 
Our results (see Table I) therefore provide further evidence 
of the  universal nature of $x_l$ for contour loops. 

\begin{figure}
  \begin{center}
    \leavevmode
    \epsfxsize=7cm
    \epsffile{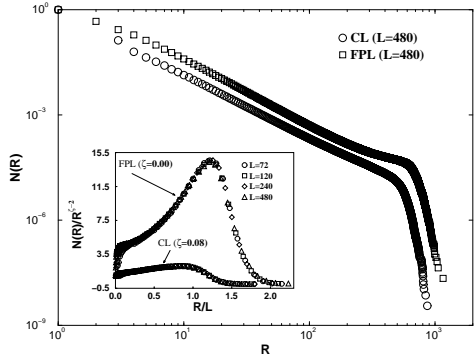}
  \end{center}
\caption{
Cumulative number density $N(R)$ (normalized by total loop number)
for both CL and FPL; $L=480$. FSS plots are shown in the inset; systematic 
errors estimated as in Fig.~\ref{fig_tau} are smaller then the 
symbols shown.
}
\label{fig_zeta}
\end{figure}

From Eq.~\ref{zeta} it follows that  the roughness exponent $\zeta$ can
be obtained from $n(R)$. Once again we  consider the
cumulative number density $N(R)$ for which we propose the scaling 
form: $N(R)=R^{\zeta-2}f_N(R/L)$. 
The results of the FSS analysis are given in Fig~\ref{fig_zeta} 
which yields $\zeta=0.08\pm 0.01$ and 
$\zeta=0.00\pm 0.01$ for CL and FPL, respectively. 

Given the numerical values of the geometrical exponents in the case of the 
{\it random elastic medium}, which are summarized
in Table I, it is straightforward to check that the
scaling relations, given by Eq.~\ref{sc_rels}, hold for both
CL and FPL. Although the small non-zero roughness $\zeta=0.08$
appears to be in disagreement with the  super-rough scenario
($\zeta = 0$), it can in fact be understood as an {\it
effective} exponent due to the extra log-divergence of the interface
width. We therefore  expect this effective roughness to approach zero 
with increasing system size as $1/\ln{L}$.

We also carried out numerical simulations for the {\it random manifold}.
Our results for the geometrical exponents are  summarized in Table I. 
They were obtained in the exact same fashion as in the case of the 
random elastic medium 
using a FSS analysis for  
$L=64$, $72$, $96$, and $150$, with the interfaces confined between two 
boundary layers separated by  $H=55$, $58$, $65$, and $78$, 
respectively; see Fig.~\ref{fig_111}. For each system size  
$6\times10^3$ disorder realizations were simulated.
Our three main conclusions are:  
First, the roughness 
exponent $\zeta=0.40(2)$ obtained from $N(R)$ 
is in excellent agreement with that obtained from a 
more traditional approach that relies on measuring height 
fluctuations\cite{GS-numerics}.  
Second, as in the case of the {\it random elastic  medium}, 
geometric exponents of both CL and FPL satisfy  scaling 
relations given in Eq.~(\ref{sc_rels}) giving strong support to the claim that 
this glassy interface is self-affine. 
Finally, and quite surprisingly, 
the geometric exponents for FPL remain unchanged from the 
random-periodic case. Moreover, their
values agree, within statistical errors, with those obtained 
in the absence of disorder, where the interface is marginally
rough due to entropic fluctuations\cite{FPL}.

In conclusion, we have measured the geometrical exponents of  
contour loops and fully packed loops associated with the ground states
of interfaces in the presence of quenched random and periodically-random 
disorder. The contour loop results are consistent with the ground state
interfaces being self-affine and rough, with roughness exponents in 
agreement with renormalization group calculations.  The geometrical 
exponents for the fully packed loops were found to be unaffected by the 
disorder. This discovery is rather unexpected in light of the fact that 
there is a 1-1 mapping between the two loop types, and it calls for a 
more detailed study of disordered fully packed loop models with 
general loop weights. 

We thank J. Cardy, C.L. Henley, J. Jacobsen, P. Leath, C. Marchetti, 
and V. Pasquier for useful discussions.
We acknowledge support by the NSF through grants 
DMR-9214943 (CZ), DMR-9357613 (JK), and 
DMR-9702242 and Alfred P. Sloan Fellowship (AAM).  

\begin{table}
\caption{Geometric exponents of both contour loops (CL) and fully-packed
loops (FPL). Rational numbers are the proposed exponents.}
\begin{tabular} {|ll|}
         Random Elastic Medium & \\ \hline\hline\hline
         Contour Loops (CL)       &Fully-Packed Loops (FPL) \\ \hline
$D     = 1.46\pm 0.01$ ($3/2$)& $D     = 1.75\pm 0.01$ ($7/4$) \\ \hline
$\tau  = 2.32\pm 0.01$ ($7/3$)& $\tau  = 2.15\pm 0.01$ ($15/7$)\\ \hline
$x_l     = 0.50\pm 0.01$ ($1/2$)& $x_l     = 0.25\pm 0.01$ ($1/4$) \\ \hline
$\zeta = 0.08\pm 0.01$ ($0$)  & $\zeta = 0.00\pm 0.01$ ($0$)   \\
\hline\hline\hline
         Random Manifold          &     \\ \hline\hline\hline
         Contour Loops (CL)       &Fully-Packed Loops (FPL)    \\ \hline
$D     = 1.31\pm 0.02$ (?)     & $D     = 1.74\pm 0.01$ ($7/4$) \\ \hline
$\tau  = 2.19\pm 0.02$ (?)     & $\tau  = 2.15\pm 0.01$ ($15/7$)\\ \hline
$x_l     = 0.49\pm 0.02$ ($1/2$)& $x_l     = 0.25\pm 0.01$ ($1/4$) \\ \hline
$\zeta = 0.40\pm 0.02$ (?)     & $\zeta = 0.01\pm 0.01$ ($0$)   \\
\end{tabular}
\end{table}

\end{multicols}
\end{document}